\documentclass[twocolumn,showpacs,amsmath,amssymb,prl,superscriptaddress,floatfix,groupedaddress]{revtex4-1}
\usepackage{graphicx}
\usepackage{bm}
\usepackage{amsmath,amsfonts}

\begin{document}

\title{Out-of-equilibrium structures in strongly interacting Rydberg gases with dissipation}

\author{Igor Lesanovsky}
\author{Juan P. Garrahan}
\affiliation{School of Physics and Astronomy, University of
Nottingham, Nottingham, NG7 2RD, UK}


\date{\today}

\begin{abstract}
The non-equilibrium dynamics of a gas of cold atoms in which Rydberg states are off-resonantly excited is studied in the presence of noise. The interplay between interaction and off-resonant excitation leads to an initial dynamics where aggregates of excited Rydberg atoms slowly nucleate and grow, eventually reaching long-lived meta-stable arrangements which then relax further on much longer timescales.  This growth dynamics is governed by an effective Master equation which permits a transparent and largely analytical understanding of the underlying physics.  By means of extensive numerical simulations we study the many-body dynamics and the correlations of the resulting non-equilibrium states in various dimensions.  Our results provide insight into the dynamical richness of strongly interacting Rydberg gases in noisy environments, and highlight the usefulness of these kind of systems for the exploration of soft-matter-type collective behaviour.
\end{abstract}

\maketitle
Ultracold gases of atoms permit the observation of many-body phenomena within a setting that offers tunable interactions together with the ability to also control the dissipative environment. Of great current interest are atoms in highly excited states, so-called Rydberg atoms \cite{Low12}, as they feature strong and long-ranged coherent forces which typically manifest in pronounced collective behavior. Recent experiments have exploited this for the exploration of dynamically prepared correlated states in lattice spin systems \cite{Viteau11,Schauss12} and the study of non-equilibrium phase transitions \cite{Carr13,Malossi13}. Furthermore, excitation transport \cite{Ditzhuijzen08,Maxwell13,Gunter13} and the formation of small aggregate structures \cite{Liebisch2005,Schwarzkopf13,Malossi13,Schempp14} were observed in driven Rydberg gases. Implicitly, all these experiments highlight a link between problems in interacting cold atomic systems and those  in (classical) soft-condensed matter, which traditionally deals with the static and dynamical collective behaviour, both in and out of equilibrium, of many-body systems with excluded volume or more complex interactions \cite{Chaikin2000,*Binder2011}.

In this work we connect to this perspective by investigating in detail the non-equilibrium dynamics of a cold atomic gas in which atoms are off-resonantly excited to high-lying Rydberg states where they strongly interact. We find an intricate sequence of distinct dynamical regimes due to the competition of interactions between excited atoms and the drive towards a high density of excitations.  The initial stages of this dynamics is one of nucleation and subsequent growth of aggregates of excited atoms, leading to long-lived meta-stable states which can display pronounced anti-correlations.  Eventually, on much longer timescales, the long-lived states relax towards true stationarity.  We analyse all these non-equilibrium features, in spatial dimensions one, two and three, by means of effective Master equations with operator-valued rates.  We use this approach to gain a largely analytical understanding of the time scales involved, and to study the correlation properties of meta-stable states by means of large scale numerical simulations.  Beyond uncovering the dynamical richness of this system our goal is to highlight the general usefulness of Rydberg gases as a platform for exploring collective phenomena traditionally associated to soft-matter physics.  We hope that this perspective stimulates new experiments exploring the role of strong interactions and quantum and thermal noise on the non-equilibrium structures of driven many-body systems.

\begin{figure}
\includegraphics[width=.9\columnwidth]{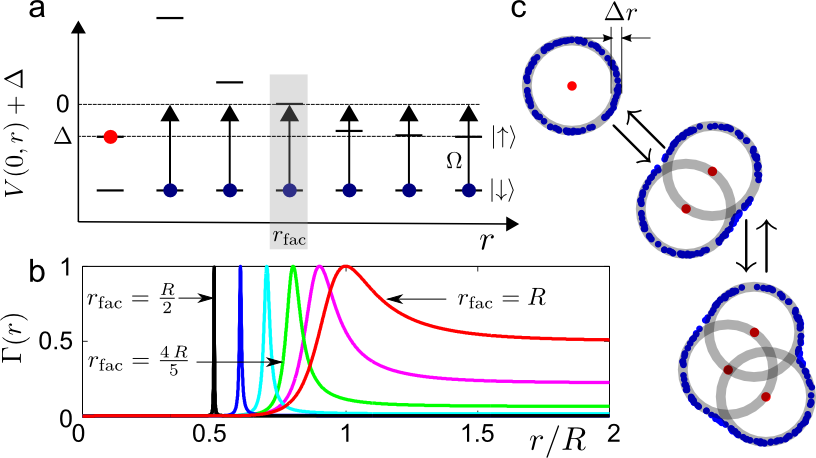}
\caption{a: Relevant atomic levels and interaction energy shifts. Atoms are excited with an off-resonant laser with detuning $\Delta<0$ and Rabi frequency $\Omega$. The interaction with an excited atom (red) shifts the transition of a second atom, positioned at the facilitation radius $r_\mathrm{fac}$ into resonance.  b: Shape of the rate function $\Gamma(r)$ for two atoms (one is excited and positioned at the origin) for $r_\mathrm{fac}=\frac{R}{2},\frac{3\,R}{5},\frac{7\,R}{10},...,R$. c: Facilitation dynamics in two dimensions. A single atom facilitates the excitation of (blue) atoms within a shell (gray region) of width $\Delta r$. Upon a subsequent excitation on this shell the facilitation region is changed to an ellipsoid. Note the deviation of the facilitation region in step two (three) from that of two (three) overlapping circles.}
\label{fig:rate_function}
\end{figure}

We follow the standard description of a frozen Rydberg gas in terms of an ensemble of $N$ two-level atoms, with internal states $\left|\uparrow\right>$ (Rydberg state) and $\left|\downarrow\right>$ (ground state). The atoms are localized at positions $\mathbf{r}_k$ which (for the sake of simplicity) are taken to be the sites of a regular $d$-dimensional lattice with spacing $a$. Excited atoms, i.e. atoms in state $\left|\uparrow\right>$, interact with a van der Waals potential of the form $V(\mathbf{r}_k,\mathbf{r}_n)=C_6/|\mathbf{r}_k-\mathbf{r}_n|^6$ which is parameterized by the dispersion coefficient $C_6$. Atoms are (de-)excited via a laser with Rabi frequency $\Omega$ and detuning $\Delta$. Furthermore, we consider the presence of noise that dephases superpositions of the atomic ground and excited state at a rate $\gamma$ as e.g. discussed in Refs. \cite{Ates06,Ates07,Honer11,Heeg12,Garttner13,Honing13,Petrosyan13-1,Lesanovsky13,Schonleber14}.
In our description we neglect the (radiative, black-body) decay of the excited state to other states which will be justified a posteriori. The many-body density matrix $\rho$ of this system evolves under the quantum Master equation $\partial_t\rho=\left(\Omega\mathcal{L}_\Omega + \Delta\mathcal{L}_\Delta + \frac{C_6}{a^6}\mathcal{L}_V+\gamma\mathcal{L}_\gamma\right)\rho$
with the Liouvillians $\mathcal{L}_\Omega=-i\sum^N_k[\sigma^x_k,\bullet]$, $\mathcal{L}_\Delta=-i\sum^N_k[n_k,\bullet]$, $\mathcal{L}_V=-(i/2)\sum^N_{k\neq m}(V(\hat{r}_k,\hat{r}_m)/C_6)[n_k\,n_m,\bullet]$ and $\mathcal{L}_\gamma=\sum^N_k(n_k\bullet n_k-(1/2)\left\{n_k,\bullet\right\})$. The operators $\sigma^i_k$ ($i=x,y,z$) are Pauli matrices acting on site $k$, and we have introduced the scaled positions $\hat{r}_k=\mathbf{r}_k/a$, as well as the projector onto the excited state of the $k$-th atom, $n_k=(1+\sigma^z_k)/2$.

In order to understand the relative importance of the individual terms we introduce the scaled time $\tau \equiv \gamma t/\alpha^2$, that depends on the dimensionless parameter $\alpha \equiv \gamma/(2\Omega)$, and which quantifies the strength of the dephasing noise relative to the coherent driving. We furthermore introduce the scaled detuning $\delta \equiv 2\Delta/(\gamma\,R^6)$, and the interaction parameter $R \equiv a^{-1} (2C_6/\gamma)^{1/6}$ that serves as a measure for the range of the interaction among Rydberg states. The Master equation acquires now the form
\begin{eqnarray}
  \partial_\tau\rho=\left[\frac{\alpha}{2}\mathcal{L}_\Omega + \frac{\alpha^2}{2} R^6\left( \delta\mathcal{L}_\Delta + \mathcal{L}_V\right)+\alpha^2\mathcal{L}_\gamma\right]\rho.\label{eq:scaled_meq}
\end{eqnarray}
This equation can be further simplified in the regime of strong dephasing noise, i.e. $\alpha>1$, by the use of second order perturbation theory in $1/\alpha$. This leads to an effective Master equation that governs the evolution of the populations in the classical basis formed by products of the $\sigma^z$-eigenstates $\left|\downarrow\right>$ and $\left|\uparrow\right>$ (see Ref. \cite{Lesanovsky13}):
\begin{eqnarray}
  \partial_t \mu =\sum_k \Gamma_k\left[\sigma_x^k \mu \sigma_x^k-\mu\right].\label{eq:effective_meq}
\end{eqnarray}
Here $\mu= P\rho$, where $P$ is a projector on the aforementioned classical basis. The effective dynamics is that of single spin flips whose rate
\begin{eqnarray}
  \Gamma_k=\frac{1}{1+\epsilon_k^2}, \label{eq:rate_func}
\end{eqnarray}
is determined by the operator
\begin{equation}
\epsilon_k \equiv R^{6}\left(\delta+\sum_{m\neq k} \frac{n_m}{|\hat{r}_k-\hat{r}_m|^6}\right) ,
\end{equation}
i.e.\ the classical interaction energy (in scaled units) that is gained/released by flipping the $k$-th spin. When this energy change is large the rate is highly suppressed, which leads to a strongly correlated dynamics of the system. The treatment of the dynamics of Rydberg gases in terms of rate equations has been employed previously also by other authors \cite{Ates06,Ates07b,Heeg12,Garttner13,Honing13,Petrosyan13-1} and it has recently been successfully used in the modelling of experimental data \cite{Schempp14}. The advantage of Eq.\ (\ref{eq:effective_meq}) is that it reduces the quantum many-body problem to a set of classical rate equations which can be simulated efficiently with Monte Carlo methods.

Let us first analyse the rate function (\ref{eq:rate_func}) and the resulting dynamics on a few-body level. To this end we consider a situation in which there is an excited atom positioned at $\hat{r}_1=0$ and all other atoms are in the ground state (see Fig. \ref{fig:rate_function}a). The flip rate $\Gamma(r)$ of an excited atom at (scaled) distance $r$ is then given by $\Gamma(r)=[1+R^{12}(\delta+r^{-6})^2]^{-1}$.  For negative detuning, this rate is optimal at a distance given by the ``facilitation radius" \cite{[{We use this terminology in analogy with kinetically constrained glass models, where the presence of excitations also alters the rate for transitions of neighbouring sites, and thus ``facilitate'' dynamics.  See for example }]Fredrickson1984,*Garrahan2002,*Ritort2003}, $r_\mathrm{fac} \equiv |\delta|^{-1/6}$, in terms of which it reads, $\Gamma(r)=[1+[(R/r)^6-(R/r_\mathrm{fac})^6]^2]^{-1}$.  The rate function is
displayed in Fig.\ \ref{fig:rate_function}b for various values of $r_\mathrm{fac}$. For small distances $r$ it is strongly suppressed before a peak emerges at $r_\mathrm{fac}$. Here the negative detuning matches the change in interaction energy caused by a single spin flip making this process resonant and therefore facilitated. For large interparticle distances $\Gamma(r)$ saturates at $\Gamma(\infty)\approx R^{-12}\delta^{-2}=(r_\mathrm{fac}/R)^{12}$. The latter expression is only valid if $R>r_\mathrm{fac}$, and only when this is satisfied the rate function gives rise to a clearly delimited peak which facilitates the excitation of atoms within a thin shell (see gray region in Fig.\ \ref{fig:rate_function}c) of width $\Delta r \approx 1/(3R^6 |\delta|^{7/6})=(r_\mathrm{fac}/3)\times (r_\mathrm{fac}/R)^6$. These considerations give rise to a qualitative understanding of the dynamics \cite{Ates06,Garttner13,Schonleber14}: From an initial state without excitations the off-resonant laser driving creates Rydberg atoms at a rate $\Gamma(\infty)$. The first Rydberg atom (see Fig. \ref{fig:rate_function}c) serves as nucleation center that facilitates the creation of excitations in its vicinity at a timescale $\sim 1$. We anticipate this to lead to the successive growth of aggregates that eventually fill up all available space at a density $\sim r^{-d}_\mathrm{fac}$. These aggregates, however, can only be transient states of the Master equation (\ref{eq:effective_meq}) whose actual stationary density matrix is given by the completely mixed state $\mu_\mathrm{ss}=\mathbb{I}/2^N$.

\begin{figure}
\includegraphics[width=0.6\columnwidth]{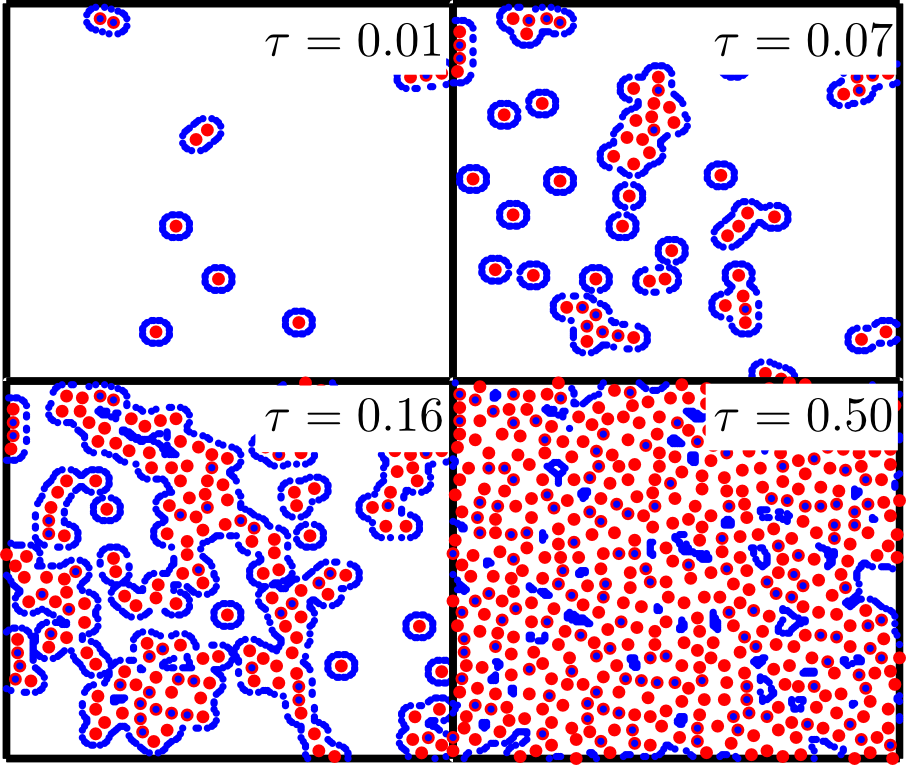}
\caption{Snapshots of a two-dimensional system with $N=200\times 200$ atoms. The interaction parameter is $R=10$ and the facilitation radius is $r_\mathrm{fac}= \frac{7\,R}{10}$. Red dots correspond to excited atoms. Blue dots correspond to atoms where state changes in the next step are facilitated, i.e. where the corresponding rate for a state change is larger than $0.1$.}
\label{fig:snapshots}
\end{figure}
To get an impression of the actual many-body dynamics we show in Fig. \ref{fig:snapshots} snapshots taken at four different times for a two-dimensional system with $R=10$ and $r_\mathrm{fac}=7$. Each excited atom is represented by a red dot. To visualize the actual excitation dynamics we mark atoms which are likely to be (de-)excited during the next Monte Carlo step with a blue dot, i.e. those atoms whose flip-rate is larger than $0.1$. One clearly observes the growth of aggregates nucleated by initial independent excitations. Eventually, excited atoms fill up the available volume reaching a meta-stable state with density $\sim r_\mathrm{fac}^{-2}$. The snapshots also show that upon nucleation aggregates grow from the boundary and that excited atoms located within an aggregate are actually ``locked in place". The reason for this effect is that the rate $\Gamma_k$, which determines the flipping rate of the $k$-th atom, depends on its interaction with \emph{all} other atoms. This immobilizes excited atoms that are located in close vicinity to two or more excitations. An example is shown in Fig. \ref{fig:rate_function}c: In the second excitation step where three atoms are excited it appears as if a flip of the left red atom was facilitated since it is located on the facilitation radius of the upper and the lower red atom. However, there is no facilitation as the actual interaction energy of the left atom is twice as large as is required for making its state change a resonant process. Beyond forming the basis of the aggregate growth such configuration dependent energy shifts have potential practical applications, for example for the realization of quantum gate protocols as discussed in Ref.\ \cite{Garttner13}.

\begin{figure}
\includegraphics[width=\columnwidth]{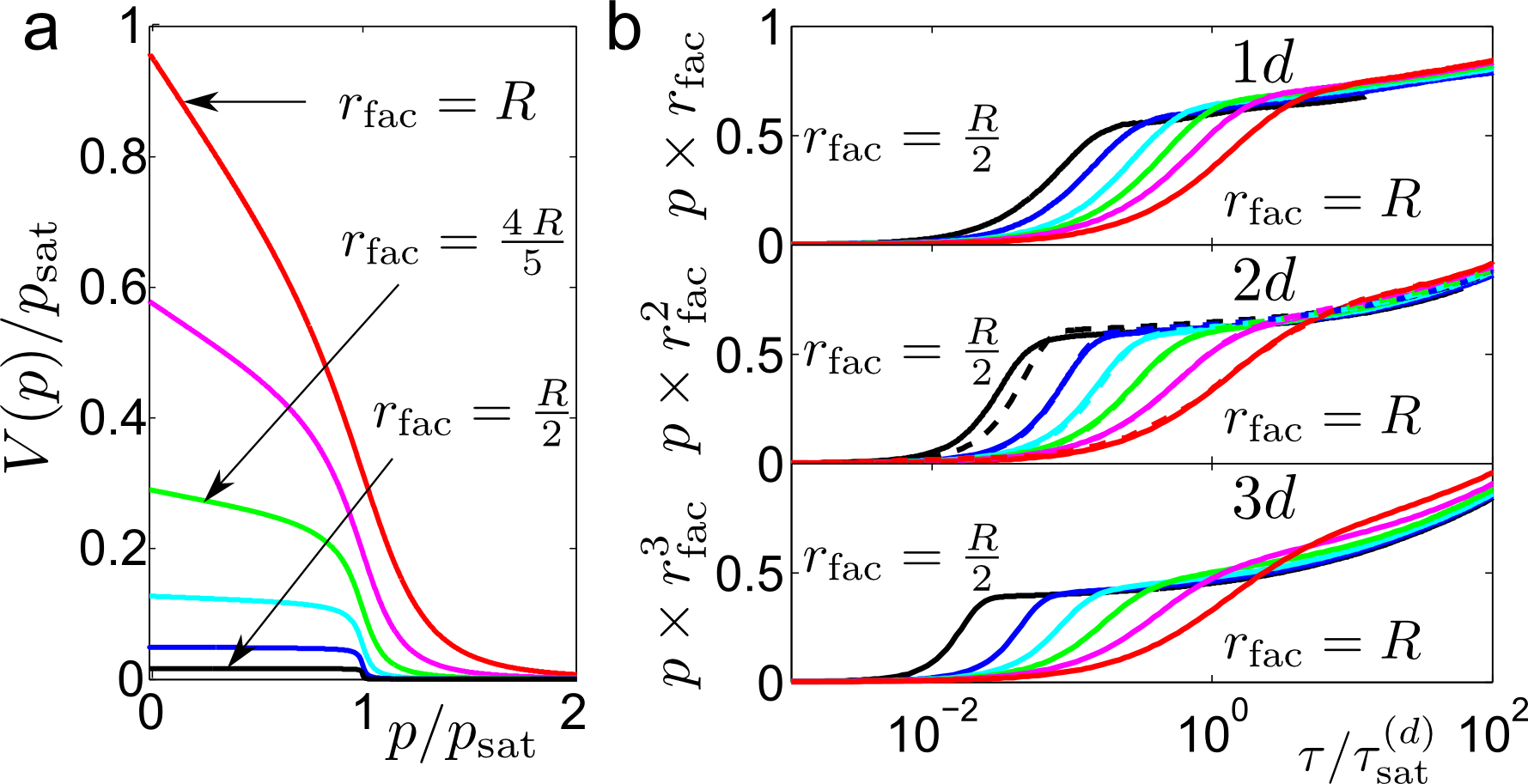}
\caption{a: Effective mean-field potential $V(p)$ [see Eq.\ (\ref{eq:mean_field})] for a two-dimensional gas and $r_\mathrm{fac}=\frac{R}{2},\frac{3\,R}{5},...,R$. b: Time evolution of the density in one ($N=10^6$), two ($N=1000\times 1000$) and three ($N=200\times 200\times 200$) dimensions. The solid lines represent data for $R=10$. Broken lines in 2d show data for $R=30$ and $r_\mathrm{fac}=\frac{R}{2},\frac{3\,R}{5},...,R$ (here $N=3000\times 3000$).}
\label{fig:density}
\end{figure}

More insight can be gained by a simple mean field analysis. To this end we consider the excitation density, $p \equiv p(\tau) = \sum_k \left<n_k(\tau)\right>/N$ and using Eq.\ (\ref{eq:effective_meq}) we calculate the equation of motion for $\left<n_k(\tau)\right>$. Assuming a homogeneous system and by breaking the emerging correlators we obtain the mean field equation $\partial_\tau p=\left<\Gamma\right>(1-2 p)$, where $\left<\Gamma\right>=[1+R^{12}(\delta+\left<\sum_m n_m/m^6\right>)^2]^{-1}$. We consider the regime where $p\ll 1/2$, i.e. $1-2p\approx 1$. Here we can obtain a very crude estimate of $\left<\Gamma\right>$ by using the fact that initially excitations get created at distances much larger than $R$ in an otherwise uncorrelated manner.  In terms of the density this distance is $l \approx p^{-1/d}$. A randomly chosen position will typically reside in a void of size $l$, and the sum $\sum_m n_m/m^6$ will be dominated by the closest excitation, which will typically be at a distance $l$, so that $\left<\sum_m n_m/m^6\right> \approx l^{-6} \approx p^{6/d}$. This yields the equation
\begin{align}
  \partial_\tau p\approx\left[1+\left(\frac{R}{r_\mathrm{fac}}\right)^{12}\!\!\left(1-\left[r_\mathrm{fac}\, p^{1/d}\right]^6\right)^2\right]^{-1}\!\!\! \equiv -\partial_p V(p).\label{eq:mean_field}
\end{align}
The right hand side is written as the gradient of an effective potential, $V(p)$, which permits a qualitative understanding of the dynamics. It is displayed in Fig.\ \ref{fig:density}a: In the case of small $r_\mathrm{fac}$ the potential has a shallow slope for small $p$.  This is followed by a sudden drop near the saturation density $p_\mathrm{sat} \approx r^{-d}_\mathrm{fac}$, indicating an accelerated increase of the excitation density which is suggestive of the facilitated formation of aggregates. Once this has taken place the dynamics becomes again slow which is indicated by the second shallow portion of $V(p)$.  We can estimate the timescale $\tau^{(d)}_\mathrm{sat}$ associated with the aggregate growth regime by integrating the mean-field equation and evaluating at $p_\mathrm{sat}$.  This the approximate result
\begin{eqnarray}
  \tau^{(d)}_\mathrm{sat}\approx\left(\frac{R}{r_\mathrm{fac}}\right)^{12}\times r_\mathrm{fac}^{-d}.\label{eq:saturation_time}
\end{eqnarray}

Let us now investigate whether these predictions are actually reflected in the numerical simulations of Eq.\ (\ref{eq:effective_meq}). In Fig. \ref{fig:density}b we plot the density of excited atoms as a function of time in one, two and three dimensions for various values of the facilitation radius $r_\mathrm{fac}$. The solid lines show data for $R=10$ where the time coordinate is scaled by $\tau^{(d)}_\mathrm{sat}$. All curves show a kink due to the sudden formation of aggregates and a (temporary) saturation at density $r^{-d}_\mathrm{fac}$. The kink becomes more pronounced the smaller $r_\mathrm{fac}$ since only here a clear distinction between the timescales of independent and facilitated excitation is present, as can be seen from the shape of the rate function in Fig.\ \ref{eq:rate_func}a. The effect is also visible in the behavior of the mean field potential $V(p)$: The larger $r_\mathrm{fac}$ the less sudden the drop at $p_\mathrm{sat}$.

Curves corresponding to the same value of $r_\mathrm{fac}$ but different dimension exhibit their relevant features at approximately the same scaled times. This confirms the dependence of the saturation time [Eq.\ (\ref{eq:saturation_time})] on $d$. In addition we show in the 2d panel data for $R=30$. Curves with the same ratio $R/r_\mathrm{fac}$ collapse. Differences for $r_\mathrm{fac}=R/2$ arise due to the fact that in the case of $R=10$ the system still ``resolves the lattice". Hence the saturation time scale appears to depend indeed only on the ratio $R/r_\mathrm{fac}$. What Eq. (\ref{eq:saturation_time}) appears to predict incorrectly, is the functional dependence of $\tau^{(d)}_\mathrm{sat}$ on this ratio. One is able to approximately collapse the curves within each panel (such that the kink in the density occurs at $\tau/\tau^{(d)}_\mathrm{sat}\approx 1$) by changing the exponent in $\tau^{(d)}_\mathrm{sat}$ from $12$ to a smaller value, typically in the range of 7 to 9. A quantitative prediction of this exponent would require a more accurate estimate of $\left<\Gamma\right>$ in Eq.\ (\ref{eq:mean_field}).

\begin{figure}
\includegraphics[width=.8\columnwidth]{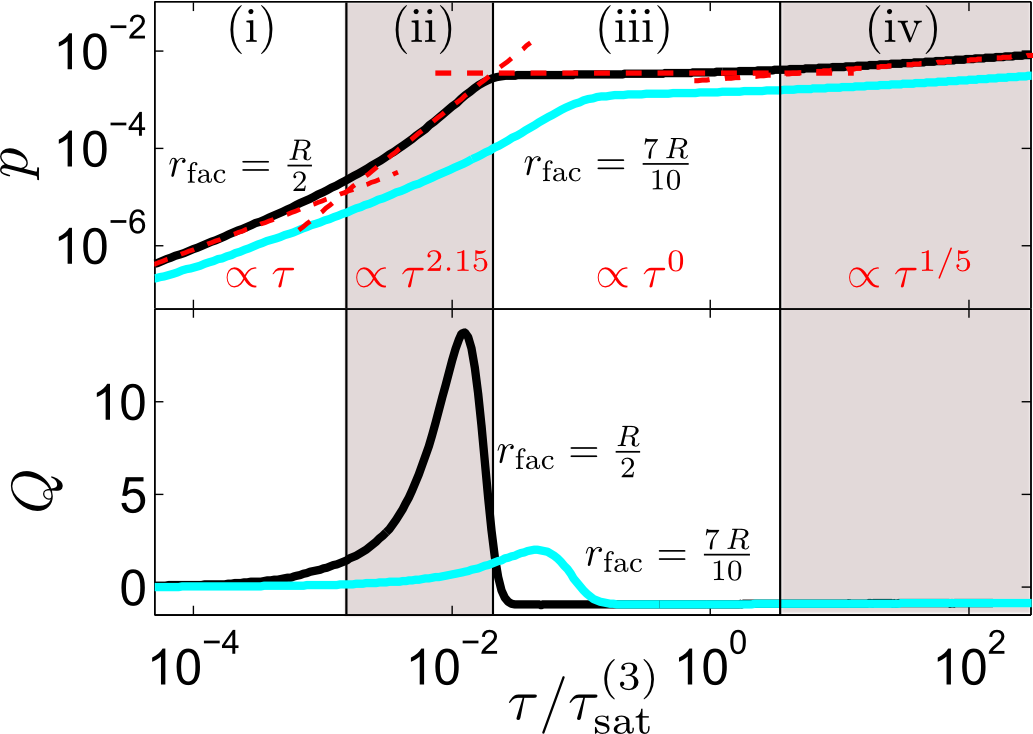}
\caption{Density and Mandel $Q$-parameter (see text) as a function of time for a three-dimensional system with $R=10$. Shown are data for $r_\mathrm{fac}=\frac{R}{2}$ ($N=30\times 30\times 30$) and $r_\mathrm{fac}= \frac{7\,R}{10}$ ($N=50\times 50\times 50$). Broken lines sketch power-law approximations to the time-dependence of the excitation density in four different regimes.}
\label{fig:scaling}
\end{figure}

The different dynamical regimes are manifested both in the scaling with time of the average density of excitations $p$ and in a change in the nature of its fluctuations.  The latter are naturally quantified via the corresponding Mandel $Q$-parameter,
$Q \equiv [ \langle (\sum_k n_k)^{2} \rangle - N^{2} p^{2}] /  (N p) - 1$,
which is commonly used in the study of interacting Rydberg gases. In Fig.\ \ref{fig:scaling} we display the excitation density on a double-logarithmic plot, together with the $Q$-parameter as a function of time. We can clearly distinguish four regimes. (i) An initial regime with a linear increase of the density $p(\tau)\approx \Gamma(\infty) \tau$, as predicted by the mean field Eq.\ (\ref{eq:mean_field}); here atoms are excited independently, which is reflected in $Q$ being close to zero. (ii) A second regime of aggregation, where $p$ increases algebraically. This can be understood from considering that aggregates are compact objects that grow from their boundary. In this case the growth rate of an aggregate with a number of excitations $n_\mathrm{agg}$ would be proportional to their surface area, $n_\mathrm{agg}^\frac{d-1}{d}$, resulting in a growth equation $\partial_\tau n_\mathrm{agg}\propto n_\mathrm{agg}^\frac{d-1}{d}$,
leading to $n_\mathrm{agg}\propto \tau^d$. The data in Fig. \ref{fig:scaling} shows a smaller exponent than the expected one for $d=3$.  We can speculate that this may be due to the fact that the boundary of the growing aggregates is less uniform than what the considerations above assume. Note, moreover, that the aggregation regime is accompanied by marked fluctuations in the number of excitations, giving rise to a large positive value of $Q$.  This is a consequence of spatial fluctuations due to the ``bunching'' of excitations in aggregates. The regime of aggregation, i.e. the algebraic increase of the density with a power different from $1$ and the dramatic increase of fluctuations, becomes less pronounced the larger $r_\mathrm{fac}/R$. (iii) After the aggregation stage the density stays approximately constant $p\sim r^{-3}_\mathrm{fac}$ for long times. Clearly, this meta-stable saturated state is strongly (anti-)correlated, as the excitations are approximately a distance $r_\mathrm{fac}$ apart from each other.  This is also reflected in the $Q$-parameter which exhibits a sharp drop to its minimum value of $-1$, indicating that fluctuations in the excitation number are highly suppressed.
Interestingly, the value of $Q \approx -1$ for these meta-stable states suggests that they might be disordered yet ``hyperuniform'' \cite{Torquato2003}.
(iv) Finally, we enter a regime in which the density, as well as the $Q$-parameter increase slowly, towards the uncorrelated $\mu_\mathrm{ss}$. Here the system increases its density by creating excitations within the gaps between already excited atoms. This leads to a hierarchical density growth with exponent $1/5$ in $d=3$, as discussed in Ref. \cite{Lesanovsky13}.

In the description of the dynamics so far we have neglected the loss of atoms from the excited states which occurs at a rate $\kappa$, either by radiative decay or the redistribution of population to other Rydberg state due to black body radiation. To show under which circumstances the processes can be neglected we estimate the probability $\pi_\mathrm{loss}$ of an atom to undergo such loss process within the saturation time $\tau^{(d)}_\mathrm{sat}$. To obtain a worst case estimate we assume that the density of excitations is saturated at $r^{-d}_\mathrm{sat}$. Multiplying the loss rate by the excitation probability and the excitation time we obtain
\begin{eqnarray}
  \pi_\mathrm{loss}<2\frac{\kappa\,\Omega}{\gamma^2}\frac{\tau^{(d)}_\mathrm{sat}}{r^{d}_\mathrm{fac}}
  =2\frac{\kappa\,\Omega}{\gamma^2}\frac{R^{12}}{r_\mathrm{fac}^{12+2d}},
\end{eqnarray}
which for instance, in the case of $\gamma=10\Omega$, $\Omega=10\kappa$, $R=10$ and $r_\mathrm{fac}=R/2$, gives rise to the probabilities $\pi_\mathrm{loss}<0.3$ (1d), $10^{-2}$ (2d), $5\times 10^{-4}$ (3d). While in general it is possible---regardless of the dimension---to drastically reduce the importance of loss, these estimates show that two and three-dimensional gases are preferable for observing the dynamical features discussed here.

In conclusion, we have studied the non-equilibrium dynamics of a driven and dissipative cold atoms gas where Rydberg states are off-resonantly excited.  The dynamics is described by an effective Master equation corresponding to individual spin-flips subject to kinetic constraints that quantify the interactions.  We found a rich behaviour of nucleation and growth of aggregates of excitations towards a highly anti-correlated meta-stable state.  These results highlight once more the connections between the collective non-equilibrium dynamics of cold atomic systems and those more traditionally associated with soft-condensed matter.

\acknowledgments
We thank O. Morsch and B. Olmos for fruitful discussions. The research leading to these results has received funding from the European Research Council
under the European Union's Seventh Framework Programme (FP/2007-2013) / ERC Grant
Agreement n. 335266 (ESCQUMA). We also acknowledge financial support from EPSRC Grant no.\ EP/I017828/1 and The Leverhulme Trust grant no.\ F/00114/BG.

\end{document}